\documentstyle[aaspp4, psfig]{article}
\newcommand{\nn}{\mbox{} \nonumber \\ \mbox{} &&}
\newcommand{\om}{\omega}
\newcommand{\ob}{\omega_B}
\newcommand{\be}{\begin{equation}}
\newcommand{\ee}{\end{equation}}
\newcommand{\ba}{\begin{eqnarray}}
\newcommand{\ea}{\end{eqnarray}}

\begin{document}

\title{ Scattering and Diffraction in Magnetospheres of Fast Pulsars}
\author{ Maxim Lyutikov}
\affil{Canadian Institute
for Theoretical Astrophysics, 60 St. George, Toronto, Ont,  M5S 3H8, Canada}
\author{Anuj Parikh}
\affil{Simon Fraser University, Department of Physics, 8888 University Drive,   Burnaby, B.C., V5A 1S6, Canada}

\begin{abstract}
We apply a theory of wave propagation through a turbulent medium to 
the scattering of radio waves in pulsar magnetospheres.
We find that 
under conditions of strong density modulation 
 the effects of magnetospheric scintillations
in diffractive and refractive regimes
may be observable. The most distinctive feature of the magnetospheric 
scintillations is their independence on frequency. 
 Results based on diffractive scattering due to small scale inhomogeneities 
 give a scattering angle that may be as large as $0.1$ radians, and a typical  decorrelation time of $10^{-8}$ seconds.
 Refractive  scattering  due to large  scale inhomogeneities
is also possible, with a typical angle of $10^{-3}$ radians and a correlation time of the
order of $10^{-4}$ seconds. Temporal variation in the plasma density may also result 
in a delay time of the order of $10^{-4}$ seconds.
The different 
 scaling of the above quantities with frequency
  may allow one to distinguish  the effects of  propagation through a  pulsar magnetosphere
from 
 the interstellar medium.
In particular, we expect that  the magnetospheric scintillations
are  relatively more  important for nearby pulsars when observed at high frequencies. 
\end{abstract}

\section{Introduction}

A number of observational results may possibly be attributed to scattering
processes inside pulsar magnetospheres. After most of  the present work was completed, the very
 interesting and convincing results of 
Sallmen et al. (1999) were published. In this work,  the authors found
{\it frequency independent} (measured at 1.4 and 0.6 GHz) jitter in the  arrival  time 
of giant pulses from the Crab (of order $100 \mu$sec). 
In addition, the frequency independence of the spread and the multiplicity of the pulse
components with  large variations in the pulse broadening times strongly
suggests that multiple components of the giant pulses are due to  refractive 
scattering inside the pulsar magnetosphere. Similarly, 
Hankins \& Moffett (1998) found that broadening times for a single
giant pulse from the Crab pulsar scale more slowly with frequency than
$\lambda ^{4}$ (the value predicted if the scattering is due solely to
the unmagnetized electron-ion interstellar plasma).
Gwinn et al. (1997) determined, using interstellar scintillation,
that the size of the Vela pulsar's radio emission region is about
500 km. This, being of order 1/10 the light cylinder radius,
is considerably larger than conventional estimates.
Similar results were obtained by Smirnova et al. (1996) and Wolszcan
\& Cordes (1987) (but see also Cordes, Weisberg \& Boriakoff 1983).
Gwinn et al. (1999) found that pulsar 0437-471 shows {\it frequency independent} 
scintillations
with bandwidth 4~MHz at several observing frequencies below 1~GHz.
Kramer et al. (1997) found that intensity fluctuations at very high
frequencies (30 GHz) are considerably enhanced if extrapolated from 
from observations at lower
frequencies. 
Kramer et al. (1999) found  frequency-dependent changes in  one  of the  pulse components of 
the msec pulsar PSR J1022+1001
   these take place on a characteristic frequency scale of 8 MHz at the frequency 1410 MHz.
With a dispersion measure of 
only 10 ${\rm \, pc/cm^3}$,  this pulsar is probably too close to show diffractive
interstellar scintillations with such bandwidth. Citing the authors of the above reference
"a propagation effect in the pulsar
  magnetosphere might be still the most probable explanation for the
  observed phenomena."

Simple models for the interstellar medium fail to fully explain the scattering
properties of nearby pulsars.
This scattering is enhanced,  relative to that extrapolated from more
distant pulsars (Sutton 1971),  and shows stronger refractive effects (Gupta et al. 1994,
Bhat et al. 1999, Rickett et al. 1999).
The proposition by Hajivasiliou (1992) and Bhat et al. (1998) that
the excess scattering of nearby pulsars lies at the surface the Local
Bubble was criticized by  Britton et al. (1998) who found strong upper limits on the angular
broadening of 5 nearby pulsars, and found that these 
sizes were inconsistent with both a uniform medium and scattering
at the surface of the Local Bubble. 
Through the examination of the angular and temporal broadening of nearby
pulsars, Britton et al. (1998) also concluded that the scattering material
of nearby pulsars lies near the pulsars, and is moving in the same
general direction as the pulsars.
Rickett et al. (1999) found that scattering material
for PSR 0809$+$74 probably lies close to the pulsar,
using a technique based on the time scale of scintillation,
although it could lie much closer if its velocity were aligned
with that of the pulsar.
Scattering in pulsar magnetospheres could contribute to  interstellar scattering
and eliminate the necessity of postulating a different,
nearby, scattering medium.
Taken together, these results compelled us to investigate the scattering of
radio waves inside the pulsar magnetosphere. 

A theory of interstellar scattering is well developed and has been
successfully applied to explain various effects observed in pulsars 
(Blandford \& Narayan 1985,Lee and Jokipii  1975a,b,c, Rickett 1977). 
In this work we will  study the 
effects of  scattering and diffraction inside the pulsar
magnetosphere.
Strong magnetic fields present in pulsar magnetospheres and the unusual electrodynamics of the 
one-dimensional electron-positron plasma both
change the 
familiar effects of scattering and refraction in plasma.  The unusual 
features of  scattering   in such plasma may allow separation from the
interstellar scattering and will serve as a tool to probe the structure
of the magnetosphere itself. 

A large variety of wave-plasma interactions, including scattering and diffraction,
may be described in terms of the variation of the refractive index of a medium.
A superstrong magnetic field plays a dominant role in defining the 
electron-photon interactions at small energies.
Typically,  the frequency of the observed radio waves is much
less that the cyclotron frequency $\om \ll \ob$. In this limit, the strong magnetic field 
suppresses the wave-plasma  interaction by $(\om / \ob)^2$ (the Thompson cross-section is smaller
by this ratio compared to an unmagnetized medium). 
For such frequencies 
the refractive index of  
strongly magnetized pair  plasma
for escaping electromagnetic modes is approximately
\begin{equation}
n^2-1\approx {  \omega_p^2 \over \omega_B^2} ,
\label{delta1}
\end{equation} 
which is fundamentally different from the case of unmagnetized electron-ion plasma
(where $ n ^2-1=  -{\omega_p^2 \over \omega^2}$). 
In Eq. (\ref{delta1}) $\omega_p$ is the plasma frequency, $\omega_B$ is the nonrelativistic 
cyclotron frequency. 
Equation (\ref{delta1}) assumes that the plasma is stationary in the pulsar frame. 
On the open field lines,  the pair
 plasma is moving with a larger streaming Lorentz factor 
 $ \gamma$. Then, the refractive index for waves propagating at comparatively
large angles to the magnetic field $\theta \gg 1/\gamma_p$ is  (see appendix \ref{modes})
\begin{equation}
n^2-1\approx { \gamma_p \omega_p^2 \over \omega_B^2} \equiv \delta.
\label{delta}
\end{equation} 
Thus, the large Lorentz factor of the moving plasma effectively  enhances the wave-plasma interaction on the open field lines.

Parameter $\delta$ is the key to the scattering and diffraction effects in the 
pulsar magnetosphere. An important  fact is that $\delta$ does  not strongly 
depend on our assumptions about the density and the streaming Lorentz factors of the 
plasma. To see this, we normalize the density of the secondary plasma to the
density of the beam, $\omega_p^2 = \lambda_M \omega_{GJ}^2 =  2 \lambda_M \omega_B \Omega$, where $\lambda_M  $ is a multiplicity factor,
and use the fact that the energy in the plasma is approximately equal to the energy in the beam,
$\gamma_p  \lambda_M \approx \gamma_b$. We find then
\be 
\delta ={ 2 \gamma_b \Omega \over \omega_B}
\ee
Theoretical  estimates (e.g. Arons 1983) of the beam's Lorentz factor give $\gamma_b \approx 10^{6-7}$.
This value is determined by the condition of vacuum breakdown due to pair production, and 
depends mostly on the structure of the magnetic field (radii of curvature) 
in the acceleration zone. In the first
approximation, $\gamma_b$ may be considered a constant for all the pulsars, regardless of their spin
period or surface magnetic field. Then for a conservative estimate  $\gamma_b \approx 5  \times \, 10^{5}$,
parameter $\delta$ near the light cylinder (where scattering is most important - see below) is
\be 
\delta = 3  \times \, 10^{-4} \left( { P \over 0.1 {\rm \, sec}}\right)^2 \,
\left({ B_{NS} \over 10^{12 } {\rm \, Gs} }\right) ^{-1}.
\label{delta2}
\ee
This is a comparatively large value. For example, a fluctuation of the order of unity in the relative density 
inside the magnetosphere  will induce time delays of the order of $\delta P \approx 30 \mu {\rm \, sec}$; 
this is a small value if compared with the total interstellar time delay for typical pulsars, but it is    
comparable to the  relative dispersive   interstellar time delay between the frequencies
separated by 1 GHz.

 Parameter $\delta $, which determines refractive
properties of the medium,  is negligible deep 
inside the pulsar magnetosphere, but 
increases with the distance from the neutron star
as $\propto r^3$. Thus, the strongest nonresonant wave-plasma interactions 
occur in the outer regions of pulsar  magnetospheres (near the light cylinder).
This  allows for  a  considerable simplification  when considering scattering 
and diffraction effects since one can  adopt a "thin screen" approximation.
We assume that emission is generated deep in the pulsar magnetosphere and then
scattered in a thin screen 
  located
near the light cylinder with a typical thickness $D \approx 0.1 R_{LC}$. 


The physical picture of the disturbances that we have in mind
consists of small scale and large scale turbulence. 
The former is excited possibly by microscopic  plasma turbulence
or nonlinear  plasma interactions 
(like 
Langmiur collapse or  modulation instability), while the latter is mostly due to the
temporal and spatial modulation of the outflowing pair plasma at the moment of its creation.
We suppose that the 
 typical sizes of the  small scale inhomogeneities  should be comparable to 
tens of Debye
radii $a_{\rm \, min}\approx 10  \times \, r_D = 10 \omega_p/v_T \approx 3  \times \, 10^2 {\rm \, cm}$
 ($v_T$ is a typical thermal velocity, $r_D$ is a Debye radius); 
while typical sizes of  the large scale  inhomogeneities are of the order
of a tenth of the light cylinder radius $a_{\rm \, max} \approx 0.1 R_{LC} = 5  \times \, 10^7 {\rm \, cm}$. 
As we will show later, the small scale inhomogeneities will contribute
to diffractive scattering, while larger scale inhomogeneities will contribute
to refractive scattering. 

Since a detailed spectrum of the density perturbations inside the pulsar magnetosphere
is  presently not known, we will also make estimates for the power law spectrum extending
from smallest  scales $a_{min}$ to largest scales $a_{max}$. 
Relating these two qualitatively different effects by a simple power law may be less
justified here than for the case of interplanetary and interstellar media.
 However, it allows for a consistent treatment of refractive and diffractive effects
and,  permits the use of well developed methods. We expect  the
results to depend qualitatively  on either the inner or the outer scales
and only weakly on the particular choice of the power law index  $\alpha $.

Another difference from the case of interstellar scattering 
worth mentioning here is that while plasma in the pulsar
magnetospheres is one dimensional, 
 we still expect the small and large scale turbulence to be
three dimensional. At large scales this is an obvious consequence of the
temporal and spatial modulation of the outflowing pair plasma;
at small scales, the  excitation of  transverse turbulence 
(modulating densities of plasma across the magnetic field)  is thought to be as
effective  as excitation of  longitudinal  turbulence (e.g., Weatherall).

Contradictory estimates exist regarding the development of 
 small scale turbulence in the pulsar plasma. The primary source
of this turbulence is usually associated with electrostatic instabilities 
which may not have enough time to develop (\cite{Melrose-DB}). 
On the other hand,
the existence of the large scale density fluctuations is certain.
If the small scale turbulence, which is responsible for the diffractive 
scattering (see below), does not develop, then  our estimates for refractive-type
effects will still be valid \footnote{ In the absence of diffractive scattering, 
refractive-type scattering may result in a formation of caustics.}.

There is 
an important unresolved problem which may affects the results of this work.
 Electromagnetic waves should be absorbed in the outer parts of the magnetosphere
 at the
cyclotron resonance:
\begin{equation}
\omega \approx {  \omega_B/ \gamma_p  }.
\end{equation}
This resonance  occurs in the outer regions at 
\begin{equation}
r \approx 3  \times \, 10^8 {\rm \, cm} \left( {\nu \over 10^9 } \right)^{-1/3}.
\left( {\gamma_p \over 100} \right)^{-1/3} \, \left( { B_{NS} \over 10^{12} } \right)^{-1/3} 
\label{qw22}
\end{equation}
Nonetheless, radiation avoids being absorbed.
\footnote{ 
The fact that we do see pulsar radio emission implies that the dipole approximation
breaks down in the outer magnetosphere.
This may be interpreted in two ways:
(1) there is a sudden decrease in the density of plasma before the 
cyclotron resonance, e.g., due to the sweepback near the leading last open field line,
(2) the field falls off slower than $1/r^3$.}

In addition, dispersion relation Eq. (\ref{delta}) is  valid only  for $\om \ll {  \omega_B/ \gamma_p  }$.
In order  to avoid the absorption at the cyclotron resonance, we limit our considerations to
fast pulsars with  periods $P$ 
\be
P < 0.13\,  {\rm \, sec} \, \left( {\nu \over 10^9 } \right)^{-1/3}
\left( {\gamma_p \over 100} \right)^{-1/3} \, \left( { B_{NS} \over 10^{12} } \right)^{-1/3}.
\ee

For numerical estimates, we will use pulsars with a period $P=0.1 \,  {\rm \, sec}$, 
surface magnetic field $B_{\rm \, NS} = 10^{12} {\rm \, G}$,
plasma frequency on the open field lines
$\om_p^2 =2\, \lambda_M\, \Omega\, \omega_B$, multiplicity factor $\lambda_M =5  \times \, 10^3 $
 and a streaming Lorentz factor $\gamma_p=100$. 
Then at the light cylinder radius $R_{LC}= 5  \times \, 10^8 {\rm \, cm}$, the
parameter for both open and closed field lines is $\delta \approx  3  \times \, 10^{-4}$.
 
\begin{figure}
\psfig{file=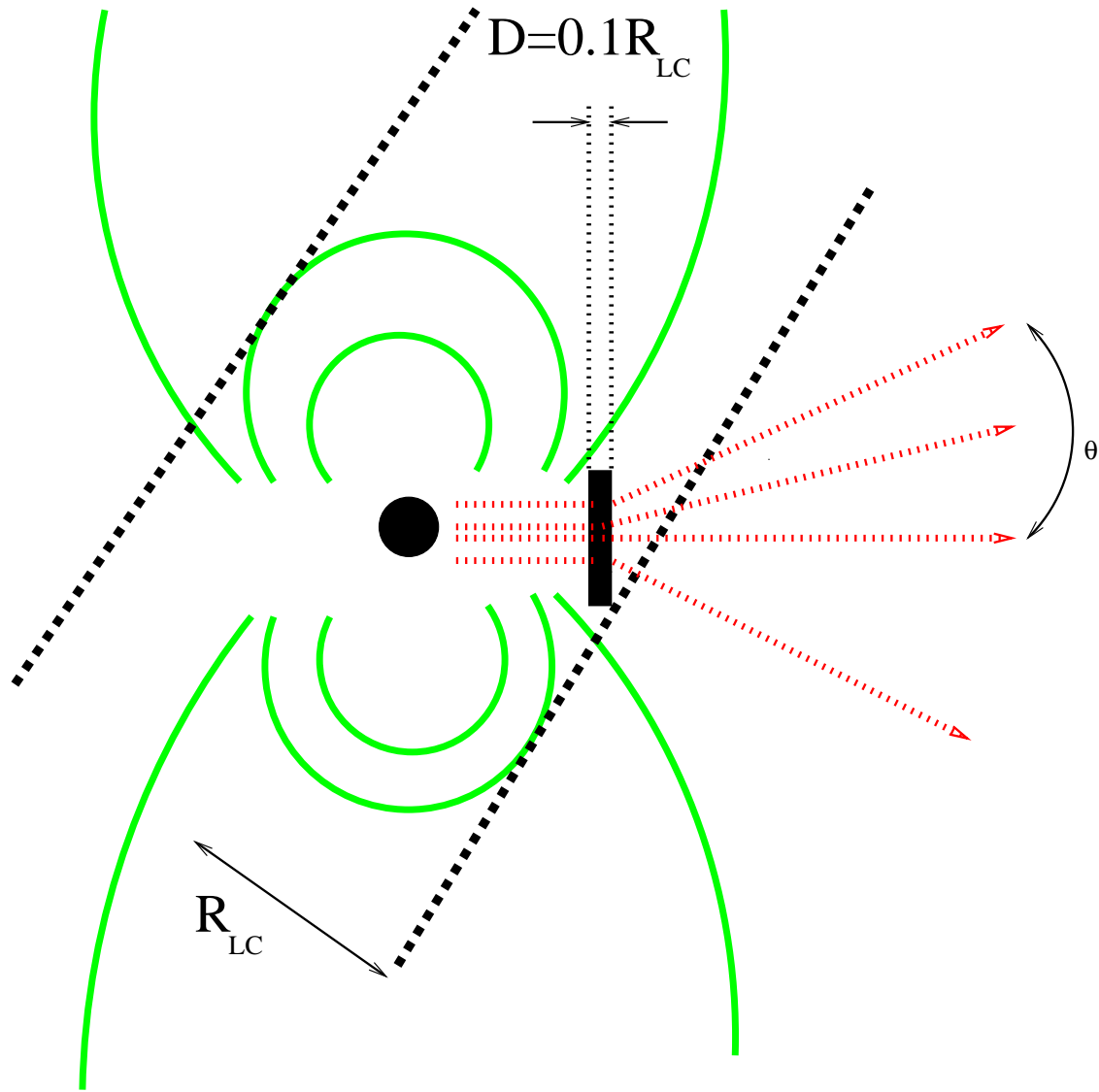,width=15.0cm}
\caption{The emission is generated deep in the pulsar magnetosphere and then
scattered in a thin screen 
  located
near the light cylinder with a typical thickness $D \approx 0.1 R_{LC}$.
}
\label{Fig1}
\end{figure}

\section{Order of magnitude estimates}

\subsection{Diffractive and refractive scattering in the case of   inhomogeneities of two scales}

Consider a screen of thickness $D$, with irregularities of
typical size $a$. A variation in the refractive index $\Delta n$
extending over a length $a$ induces a change in the phase of a wave by
$\delta \phi \approx \omega a \Delta n /c $.
 Using dispersion relation Eq. (\ref{delta}) the variation in the refractive index
may be written as
\be 
 \Delta n \approx 4 \pi r_e c^2  \gamma_p \Delta N_e /  \omega_B^2 = \delta
\Delta n_e
\ee
where $r_e =e^2 /m c^2$ is the
classical radius of an electron, and we have normalized  density perturbations as 
$  \Delta   n_e \equiv  \Delta N_e /N_e$.
 The phase shift relative to a vacuum-propagating wave after propagating the distance $a$, is 
\be
\delta \phi \approx  2 \pi  \delta \, \Delta n_e \, {a\over \lambda }
\propto \lambda ^{-1}.
\ee
The inverse wavelength dependence of the phase shift highlights the important 
point: scattering becomes stronger at high frequencies.  This is opposite  
to the scattering trend by ISM.

A ray passing through the whole screen encounters $D/a$ randomly distributed irregularities; therefore, the difference between the 
phases of rays separated laterally by more than $a$
is
\begin{equation}
\Delta \phi \approx  2 \pi \left(  {D\over a} \right) ^{1/2} \delta \phi  = 
 \left(  {D\over a} \right) ^{1/2} \delta \, \Delta n_e \, {a\over \lambda }.
\label{Deltaphi}
\end{equation}

The form of the scattered image will depend on whether the scattering is weak  (when $ \delta \phi \ll \pi$)
or strong (when $ \delta \phi \gg \pi$). 
The angular  spectrum of waves has two limiting cases:
\ba
&&
\theta_s = {\lambda \over  a}, \mbox{  $ \Delta \phi \ll \pi$}
\nn
\theta_s = { \Delta \phi \over 2 \pi} {  \lambda \over  a },   \mbox{  $ \Delta \phi \gg \pi$ }
\ea
In the case of  weak scattering, we see  the initial wave  and a thin  halo
 around the mostly unscattered wave. In this limit,
 the transverse dimension
of the correlation function, $\rho_c = \lambda /\theta_c$,
 is equal to the  size of inhomogeneities. In the case of strong scattering,
most of the energy of the initial wave has been transformed into a scattered component. 
The transverse dimension
of the correlation function  becomes smaller:
$\rho _c \approx  a/\Delta \phi $.

In the case of scattering inside the pulsar magnetosphere, the typical phase delay $\Delta \phi $  is larger
than $\pi$ (see below) - this implies that scattering occurs in the strong regime.
The rays are typically  scattered by the angle 
\begin{equation}
\theta_{\rm \, scat} \approx { \Delta \phi \over 2 \pi} {  \lambda \over  a }=
\left(  {D\over a} \right) ^{1/2} \delta \,
 \Delta   n_e.
\label{thetascat}
\end{equation}

The two types  of inhomogeneities that should be present inside the pulsar magnetosphere
will produce qualitatively different effects:  small scale inhomogeneities will produce 
diffractive scattering, while large scale inhomogeneities will produce  refractive  scattering.
To see this, we recall  (e.g. Prokhorov et al. 1975) that
strong scattering may occur in  refractive,  diffractive or mixed regimes.
Separation into refractive and diffractive effects is based on the observation
that the
geometric optics limit becomes unsuitable for distances  larger than
\be
L_{\rm \, cr}= {a_{\rm \, min} ^2 \over \lambda}.
\label{d}
\ee
For larger distances, the diffractive pattern due to the
refractive inhomogeneities becomes important regardless of the small size of the
diffraction angle.
Thus we
 can identify two scattering regimes (i) diffractive, where the inhomogeneities 
are seen mostly in the Fraunhofer zone and (ii) refractive, where   the inhomogeneities
produce fluctuations in the index of refraction and can be described by the geometrical optics
approximation. The two limits are separated by the size  of the first  Fresnel zone, 
 $r_f = \sqrt{ \lambda  R_ {\rm \, LC}  } \approx   10^5 {\rm \, cm}$.
Inhomogeneities with a size less than the size of the first Fresnel zone
contribute to diffractive scintillations, while those with a size  larger than the  first Fresnel zone
contribute to refractive  scintillations. 

For diffractive scattering, 
Eq. (\ref{thetascat}) is an estimate of the 
 the apparent angular size of the source seen through the screen.
Numerically for $D \approx 0.1 R_{LC}$,   $a = a_{\rm \, min}$ 
and  under the most optimistic assumption of
relative density fluctuations of the order of unity $\Delta   n_e \approx 1$,  we find:
\begin{equation}
 \theta_{\rm \, D} \approx 
0.1 {\rm rad}. 
\label{thetadiff}
\end{equation}
This implies that large angle scattering is possible in the outer regions
of pulsar magnetospheres.
The observed profiles  are  then the convolution of the "initial" window function 
(determined by the emission conditions at lower radii)
with  diffractive scattering.

We would like to reiterate here the the estimate of the scattering  angle 
 in Eq. (\ref{thetadiff})assumes that all
the power in the density fluctuations is concentrated at one diffractive scale and that the
 turbulence is strong $<\Delta N_e> \approx N_e$. Given the total power in 
fluctuation (e.g. $<\Delta n_e>^2 \approx 1$), the 
assumption of a power law spectrum of density fluctuations removes
some power from small scales  and reduces the diffractive scattering angle (see Sec. 2.2).
On the other hand, parameter $\delta$ increases with period of the pulsar - in the longer
period pulsars it will be larger than we assumed. 

The diffractive image will be focused and defocused by the refractive fluctuations with
a scale $\approx \theta_D R_{LC} \approx a_{max}$. A typical 
 refractive scattering angle is 
\begin{equation}
\theta_{\rm \, R} \approx \delta = 4  \times \, 10^{-3}.
\label{thetaref}
\end{equation}

Refractive effects will induce "jitter" in the arrival times of the pulses and 
a temporal 
correlation in the intensities with a
typical scale $\tau_R = \theta_{\rm \, R} P = 4  \times \, 10^{-4} {\rm \,  sec}$.
Both of these effects will be independent of frequency,
and increasing with the period of the pulsar.

As we have already mentioned,  a 
 necessary requirement to allow the separation of scintillations into diffractive and
refractive branches is that the scattering  should be  strong (the
total phase shift (Eq. (\ref{Deltaphi})) is much larger than $\pi$).
 Using  our simple estimates (Eq. (\ref{Deltaphi})) 
for a medium with given size inhomogeneities, we
  find that this condition is satisfied for
\be
a_{\rm \, min}  > { \lambda^2  \over D  \delta^2} = 2  \times \, 10^2 {\rm \, cm}.
\label{amin}
\end{equation}
Alternatively, for a given size of inhomogeneities $a > a_{\rm \, min} $ the scattering is strong
for
\be
\lambda <  \lambda {\rm \, max}  = \sqrt{ a D} \delta \approx 
\left\{ \begin{array}{ll}
30 \, {\rm \, cm} & \mbox{ for $a  = a_{\rm \, min} $}\\
10^5\,  {\rm \, cm} & \mbox{ for $a  = a_{\rm \, max} $}
\end{array} \right.
\ee
A very important point is that this is a limitation
of the wavelength {\it from above}:  scattering is stronger for shorter wavelengths.
This is in sharp contrast to the scattering in the interstellar medium, where the
strength of the scattering increases at low frequencies.

For the typical wavelength of observations $\lambda = 30 {\rm \, cm}$,
both refractive and diffractive scattering occurs in a strong scattering regime
with the  total phase change  (Eq. \ref{Deltaphi}):
\be 
\Delta \phi \approx  \left\{
\begin{array}{ll}
1 & \mbox{ diffractive}\\
100 & \mbox{ refractive}
\end{array} 
\right.
\ee
Diffractive scattering is, though, only mildly strong.
In comparison, 
in the ISM the small scale scintillations are saturated, while large scale are not.

  The diffractive  scattering angle $\theta_D$ will determine the
lateral 
size $\sigma$ of the  scatter-broadened spot:
\be
\sigma \approx \theta_D R_{LC} \approx 5  \times \, 10^7 {\rm \, cm}
\ee
Interestingly, this is very similar to the size of the emission region
of the Vela pulsar found using interstellar scintillations (Gwinn et al.1997)
and consistent with other measurements of the emission sizes 
(Smirnova et al. 1996).

Propagation of radiation through scattering media will also  introduce temporal smearing
of the pulse and related frequency decorrelation.
Usually, these effects  are 
 due to the combination of three independent contributions (\cite{LJII}): (i)
dispersive decorrelation effects arising from the nonzero
$\partial^2 \om / \partial k^2$ (which for sufficiently 
narrow receiver bandwidth effectively lead to  a decrease in  decorrelation bandwidth)
(ii) diffractive  decorrelation due to small  scale scattering and
(iii) refractive decorrelation due to the different transit times along the 
different ray paths.
An important difference in our case is that dispersive decorrelation effects
will be absent, since the waves are nondispersive.

The extra path length introduced by strong diffractive scattering 
is  $1/2 \theta_{\rm \, scat} ^2 R_{LC}$. This 
will contribute to time delay and pulse broadening 
of the order
\be
\tau _D  = 
 {  R_{LC}  \theta_{\rm D}^2 \over 2 c } = 
\approx  10^{-4} {\rm \, sec},
\label{qqq}
\ee
 (with corresponding decorrelation bandwidth 
$ \Delta \nu_D  = 2 \times 10 ^3 {\rm \, Hz}$).
There is also a comparable   group delay due to changing group velocity
\begin{equation}
\Delta t_{group}\approx   \delta  \frac{D}{c} \approx 10^{-5} {\rm \, sec}. 
\ee

In addition, the motion of the
 scatter-broadened spot   due to both the rotation of the pulsar and the motion
of scattering material inside the pulsar magnetosphere
will produce refractive-type
focusing and defocusing.
The size of the diffraction pattern  at Earth will be 
  $b \approx \lambda / \theta_{diff} \approx 10 ^{15} {\rm \, cm}$, where
\be 
\theta_{diff} \approx \theta_D {R_{LC} \over L} \approx 10^{-13} 
\ee
 is the visible size of the
scattering spot ($L$ is the distance of a pulsar to  the Earth). The motion of this diffraction pattern due to the rotation
of the pulsar will produce variations on  a diffractive scintillations decorrelation  time scale of 
\be 
\tau_ {diff} = {b\over L} P  \approx {\lambda \over c \theta_D} \approx 10^{-8} {\rm \, sec},
\ee
with associated decorrelation bandwidth
\be
\Delta \nu_{diff} = {1 \over 2 \pi \tau_{diff}} \approx  10^{7} {\rm \, Hz}.
\label{decorband}
\ee
Motion of the scattering material inside the pulsar magnetosphere will produce
diffractive scintillations on the same time scale.

In addition  the 
motion of the scattering disk of the size $\sigma$  will produce refractive
type variations with a typical time scale 
\be 
\tau_R \approx \theta_R P =  10^{-4} {\rm \, sec}
\ee
and refractive decorrelation bandwidth
\be
\Delta \nu_{R}   \approx 2 \times 10^3 {\rm Hz}
\label{nuR}
\ee

\subsection{ Motion of plasma}

Plasma in the pulsar magnetosphere is moving  along the magnetic field 
lines with relativistic velocities. This 
motion will affect the observed intensities in two
ways: (i) motion of the inhomogeneities will produce temporal variations in the observed
flux, (ii) we may expect
to see aberration effects like  relativistic contraction of the density inhomogeneities along
the direction of motion which will produce a diffractive picture that is not circularly. The 
observed scattered  image of the point source will have an elongated form
with a ratio of axis $\approx \gamma_{\perp}$ ($\gamma_{\perp}$ 
is a Lorentz factor of the motion in the plane of the sky - 
the parallel component of the perturbation's wave vector does not affect 
the scattering. For matter moving at an angle of about 45 degrees to the line of sight,
we can estimate $\gamma_{\perp} \approx \sqrt{\gamma} =10$. 
Thus, the image should be strongly elongated in the direction perpendicular to the
direction of motion of matter. 
This effect, if resolved, will help to determine an absolute position of the
rotation axis of the pulsar on the sky. 

To see how abberations appears in the general approach we note that
for the case of frozen inhomogeneities,
both motion of the plasma and relativistic contruction
 can be treated in the same systematic approach if we 
transform the distribution function of the index fluctuations  from the plasma
rest frame  $ P_k \delta(\omega)$ to the laboratory frame.

The Lorentz transformation of the power spectrum of density inhomogeneities $P^l$, is related to the untransformed power spectrum $P$ as
\begin{equation}
P^l({k^l}_{\parallel},k_{\perp}) = P( \gamma  k_{\parallel},k_{\perp}).
\end{equation}
while the  transformed correlation function of the power spectrum $\psi ^l$ is related to the untransformed correlation function $\psi$ as
\begin{equation}
\psi ^l({r^l}_{\parallel},r_{\perp}) = \psi({r_{\parallel} \over \gamma}, r_{\perp}).
\end{equation}
(here the sign ${\parallel}$ denotes componets along the velocity of the medium).
Using these relations it is easy to see that 
the relativistically moving medium is similar to  a medium with anisotropic turbulence.

\subsection{Power-law distribution of inhomogeneities}

In turbulent media, it is common for the  spectrum  of the density perturbation
to be a power law. Below  we give order-of-magnitude estimates for this case
(assuming that power law index $\alpha $ is less than four).

There are some substantial differences between scattering by a 
screen with one scale and a power law. First, the correlation function of intensity fluctuations 
 for the one-scale screen
 has scales of the order
of the size of the inhomogeneities of the phase screen;
  while  for the
power law, the correlation function of intensity fluctuations has scales of the order
of the Fresnel zone $\sqrt{\lambda R_{LC}} $. Secondly,
  the one-size screen can give strong focusing in the limit of scintillations (e.g., caustics can form if 
no diffractive size inhomogeneities are present); 
while in the case of the power-law screen,   scintillation only approaches unity.
 
The power law of density perturbations may extend from smallest scales
$a_{min}  = 1/q_1 \approx 5  \times \, 10^2 {\rm \, cm}$ to largest scales
$a_{max} =1/q_0 \approx 5  \times \, 10^7 {\rm \, cm}$.
To estimate the scattering for the power law spectrum of density perturbations
$\Psi_{N_e} \propto k^{-\alpha}$, we note that the typical mean square density fluctuation
due to  waves with $k\approx 1/ a$ is $\delta N_e^2 \approx N_e^2 ( q_0 /k)^{\alpha-3}$.
The total phase delay due to the fluctuations on the scale $a  \approx 1/k$ for
strong turbulence $ \delta N_e^2 / N_e^2 \approx 1$
is then 
\begin{equation}
\Delta \phi = \sqrt{ { D\over a} } \sqrt{ \delta n^2 } \delta { a \over \lambda} \approx
D^{1/2}  q_0 ^{(\alpha-3)/2} a ^{ {\alpha -2 \over 2} } \lambda^{-1} \delta.
\end{equation}
The mean scattering angle 
\begin{equation}
\theta = \Delta \phi  {\lambda  \over a } =  \sqrt{ { D\over a} } \sqrt{ \delta n^2 } \delta  \approx
D^{1/2}  q_0 ^{(\alpha-3)/2} \delta
a  ^{ {\alpha -4\over 2} } 
\label{meanscat}
\end{equation}
is independent of frequency.
For $\alpha < 4 $ (which includes the Kolmogorov 
spectrum $\alpha = 11/3$), the scattering is mostly due to the small scale density fluctuations;
while for $\alpha >  4 $, scattering is mostly due to the large scale refractive density fluctuations.
The mean scattering angle (Eq.\ref{meanscat}) generally has only a weak dependence on the size ($ a  ^{-1/6}$ 
for the Kolmogorov spectrum). This suggests that the contribution from all scales may not  be negligible.

Estimates  of the strength of the scintillations 
can be made for a power law distribution.
The minimum scale that contributes to the strong scattering  at a given
frequency is (assuming  $\alpha >2$)
\begin{equation}
a_{\rm \, strong} = {1\over q_0} \left( {  q_0 \over D } \right) ^{ { 1 \over \alpha -2 }}
  \lambda^{ {2  \over \alpha -2 }} \approx 2 {\rm \, cm}.
\end{equation}
Moreover, for a given size $a$ the  scattering is strong for
\begin{equation}
 \lambda <  \lambda {\rm \, max}= \sqrt{ {D \over q_0}} 
\left( q_0    a  \right) ^{ {\alpha -2 \over 2} }  \approx 20 a ^{5\over 6}  {\rm \, cm},
\end{equation}
where we assumed $q_0  \approx 1/D$ and $ \alpha =11/3$.

\subsection{Finite size of the emission region}

Untill now
we   neglected the finite size of the source 
in assuming  that the initial unscattered wave corresponds to a plane wave.
Finite size of the emission region may quench the scintillations.

The criterion for scintillation to be quenched depends on whether the
scintillation is weak or strong and whether the scintillation is
refractive or diffractive.
Weak scintillation, or strong refractive scintillation, will take
place if the angular size of the source seen from the observer is less
than the angular size of the scattering disk.  For scattering very
near the source (as magnetospheric scattering for a pulsar),
scattering will take place if the linear size of the source is less
than the linear size of the scattering disk.
Strong diffractive scintillation will take place if the linear size of
the source is smaller than the linear resolution of the scattering
disk, seen as a lens.  This can be expressed by requiring that the
size of the source be less than $\lambda/\theta_S$, where $\theta_S$
is the angular size of the scattering disk, seen from the source.  For
scattering at the light cylinder right near the source, $\theta_S$ is
the same as the total scattering angle
$\theta_D$. Note that even if diffractive scintillation is quenched by finite
size ofthe source, diffractive scattering will still lead to angular
broadening.

For a 0.1-sec pulsar, the light cylinder radius is $R_L=5 \times \, 10^8$~cm,
 the emission region size of the order of the
 polar cap  radius $R_{PC} \sim 1$~km 
refractive scintillation will take place if the scattering angle
is greater than $\Delta W \approx 2 R_{PC}/R_L \approx 5 \times \, 10^{-4}$~rad.
Diffractive scintillation will take place if the scattering angle
is {\it less} than $\lambda/(2 R_{PC})$.  For 30 cm-wavelength observations
of a 1-km polar cap, the required   scattering angle of $3  \times \, 10^{-4}$~rad.

For a msec pulsar, with the similar  emission region
size and cap and the light cylinder radius is $R_L=3  \times \, 10^7 $~cm 
($P=$5-msec)
the refractive scintillation will take place if the
scattering angle is greater than $10^{-2}$~rad
while the limits on 
diffractive scattering will  be the same as for 0.1-sec pulsar.

Quenching by finite source size  may 
 also serve as an additional test 
to distinguish interstellar and magnetospheric  diffraction: the latter should be more prominent
in  pulsars with narrow pulses.

\subsection{Millisecond Pulsars}

In this work we limited ourselves to fast rotating pulsars, for which 
the refractive index for radio wave propagation is independent of frequency. 
In  slower rotating pulsars, the magnetic field near the light cylinder (adjusted for the
relativistic motion of plasma) falls below the wave frequency; 
the plasma modes then have a 
  "normal" dispersion (refractive index $ \propto 1- \om_p^2 /\om^2$).
Scattering effects by such a plasma 
may change the simple frequency
dependence (or independence) of the predicted effects.  

 We expect that our order-of-magnitude  estimates of scattering   effects will hold for
millisecond pulsars as well. Near the light cylinder, the parameter $\delta$ 
(Eq. \ref{delta2}) is of comparable order for both millisecond pulsars and normal pulsars. 
The width of the scattering screen, which in millisecond pulsars occupies the
whole magnetosphere, is also of the same order $\approx 5  \times \, 10^7 {\rm \, cm}$.
The Fresnel radius is slightly smaller: $r_f =\sqrt{R_{LC} \lambda} \approx 4  \times \, 10^4 
{\rm \, cm}$. 
Quenching of diffractive scintillation may be more of a problem for millisecond
pulsars, however,  since they have larger emission beams.

\section{General considerations}

In this section we give general formulae for the scattering of electromagnetic waves  
by turbulent media within the pulsar magnetosphere. If the smallest size of the inhomogeneities
is much larger than the wavelength, we can use the scalar wave equation to describe the
wave scattering. 
Using the size of the first Fresnel zone $r _F =\sqrt{ \lambda R_{\rm \, LC} } \approx 10^5 {\rm \, cm}$ for comparison purposes,
we see that scattering by the inhomogeneities larger than $r _F$ occur in the geometric optics
limit, while scattering by the inhomogeneities smaller than $r _F$ occur in the Fraunhofer limit.
To avoid  considering two models of scattering (as did Blandford \& Narayan)
we employ the Markov approximation, which is valid for short wavelengths satisfying both
i) $\lambda \ll a_{\min}$ ($a_{\min}$ is the smallest scale of inhomogeneities)
and
ii) $\lambda \ll l_c$ where $l_c$  is the radius of coherence of  the field (see below) (Rytov et al.).

In the Markov approximation, the  medium is assumed to be a collection of uncorrelated scattering 
phase screens - the correlation function of the refractive index fluctuations
$\psi({\bf r})$ is assumed to be delta correlated along the line of propagation:
\be
\psi_n({\bf r}) = < u({\bf r}_{\perp},z)  u^{\ast}({\bf r}_{\perp}',z')>=
 A({\bf r}_{\perp}-{\bf r}_{\perp}') \delta(z-z')
\ee
(Rytov et al.).
Then 
 $A({\bf r}_{\perp}) $ - the two dimensional correlation function - is related to the
refractive index fluctuations as
\be
A({\bf r}_{\perp}) = \int dz \psi_n({\bf r}_{\perp},z) =
  \int J_0( k_ {\perp} r_{\perp})  P_n({\bf k}_{\perp},0) { d^2 {\bf k}_{\perp} \over (2 \pi)^2 }.
\ee
The physical meaning of the correlation function is that
$ z = { \lambda ^2 \over   A(0)}$ gives a length at which most of the energy of the
wave is converted from ordered component into fluctuating.
Related to the correlation function is the 
 two dimensional structure function of the  refractive index fluctuations for a homogeneous medium:
\be
D({\bf r}_{\perp}) = 2 \left( A(0) -A({\bf r}_{\perp})  \right) =
{1\over \pi} 
\int _0^{\infty}  \left(1 -J_0( k_ {\perp} r_{\perp}) \right) P_n (k_ {\perp},0) k_{\perp}dk_{\perp}. 
\ee

The scattering properties  of the pulsar magnetosphere are well
approximated by that of an
equivalent screen.
The equivalent screen  structure function  of the phase fluctuations for a homogeneous medium  is
\be
D_S({\bf r}_{\perp})= { D \over \lambda ^2} D({\bf r}_{\perp})=
{8 \pi^2 r_e^2 r_L^4 D  \gamma_p^2  \over \lambda^2} \int _{q_1}^{q_0} d k_{\perp} k_{\perp}
\left( 1- J_0( r_{\perp} k_{\perp}) \right) \Psi_{\Delta N_e} (k_{perp}),
\ee
where $r_L = c/\omega_B$ is the Larmor radius.

The  normalized 
correlation function of the wave field $u( {\bf r}_{\perp},z)$  immediately after the screen is 
\be
\Gamma(  r_{\perp},z) = \exp\{ - D({\bf r}_{\perp})_S\} =
 < u( {\bf r}_{\perp},z) u^{\ast} ( {\bf r}_{\perp},z)> =
\exp \{ - { z \over \lambda ^2}  D({\bf r}_{\perp}) \}.
\ee

One can define several  scales associated with the phase structure
function $D({\bf r}_{\perp})_S$:
(i) the typical scale of $D(\rho_c)_S = 1$ determines the transverse coherence radius
$\rho_c$, so that
the fields separated laterally by less that $\rho_c$ are
strongly correlated; (ii)
a longitudinal length  $z_c = {\lambda ^2 \over  D({\bf r}_{\perp}) }$ which
determines the minimal distance at which the Markov approximation is applicable;
(iii)
second (and larger) derivatives at zero displacement determine the 
typical small scale ($ r_{\perp} \ll a_{\rm \, min}$) behavior of the
phase structure function
\be
a_2^{-2} ={\partial^2 D(  0)_S \over \partial  r _{\perp}^2}.
\ee

For larger scale inhomogeneities (such that $a_{\rm \, min} \gg \lambda$), the transverse correlation radius
after the screen
is equal to the transverse correlation radius on the screen $l_{\perp}=\rho_c$; while  
the longitudinal correlation  length is  $l_{\parallel} = \rho_c^2 /\lambda \gg \rho_c$.

\subsection{Power law distribution}

We consider the power spectrum of the  density  perturbations $\Psi_{N_e} ({\bf k})$ 
as the initial quantity for
our approach. It is related to the  the power spectrum of the   refractive index perturbations $P_n({\bf k})$ as
\be
P_n({\bf k}) = \delta^2 { \Psi_{N_e} ({\bf k}) \over N_e^2}.
\ee
We will use the following approximation for the
 refractive index perturbations:
\begin{equation}
P_n({\bf k}) = {B \exp\{ - k^2/q_1^2\} \over (1 +k^2/q_0^2)^{\alpha/2} },
\label{powspec}
\end{equation}
where 
the outer scale is $1/q_0 = a_{\rm \, max} \approx  D$, and the inner scale is $1/q_1 = a_{\rm \, min}$.
The dependence of the various effects on the power law index is complicated. In what
follows we assume that $ 3 < \alpha < 4$ (which includes the Kolmogorov value of
$\alpha =11/3$).

The correlation function of the refractive index perturbations
is
\be
\psi (r) = \int e^{i {\bf k \cdot r}} P_{q}d^{3}q
\approx 
\left\{
\begin{array}{ll}
(q_{0}r)^{\alpha }r^{-3}B = (q_{0}r)^{\alpha -3}\delta ^{2}
 & \mbox{ $ \frac{1}{q_{1}}<r<\frac{1}{q_{0}} $ } \\
(\frac{q_{0}}{q})^{\alpha } B  =
q_{1}^{5-\alpha }q_{0}^{\alpha -3}r^{2}\delta ^{2} & \mbox{
 $ r\ll \frac{1}{q_{1}} $}
\end{array} 
\right.
\ee

The normalization of $P_n$ is determined by the the mean square of the refractive index perturbations:
\be
\left\langle \varepsilon ^{2}\right\rangle = \int d^3 {\bf k} P_n 
\ee

\be 
B=\frac{8\pi ^{3/2}\Gamma (\frac{\alpha }{2})}{q_{0}^{3}\Gamma (\frac{\alpha }{2}-\frac{3}{2})}\delta ^{2}
\left( \frac{\left\langle \delta N_{e}^{2}\right\rangle }{N_{e}^{2}}\right)
\ee
To normalize the spectrum of the refractive index perturbations, we assume that the turbulence
is strong, i.e., the mean square of the relative density perturbations is of the order
of unity:
\be
{ \left\langle \delta N_{e}^{2}\right\rangle \over  N_{e}^{2} } = 1.
\ee
Then, we obtain the following expression for the coefficient B:
\be
B=\frac{8\pi ^{3/2}\Gamma (\frac{\alpha }{2})}{q_{0}^{3}\Gamma (\frac{\alpha }{2}-\frac{3}{2})}\delta ^{2}\sim 1.3 \times \, 10^{18}{\rm \, cm}^{3}. 
\label{coeffB}
\ee
For \( \alpha =11/3, \) \( q_{0}=10^{-8}cm^{-1}, \) and \( \delta =3 \times \, 10^{-4}, \)

Note, that for the chosen range of $\alpha$, the normalization of the turbulent
spectrum depends only on the outer scale; in this case, the increase of the power 
spectrum at small wave vectors outweighs the large phase volume at large wave vectors.

 For the power spectrum (Eq. \ref{powspec}) for distances much larger than the inner scale 
 $r _{\perp} \gg 1 /q_1$, 
the 2-D correlation function is
\be
A({\bf r}_{\perp}) = { B q_0^2 ( q_0  r _{\perp})^{\mu} K_{\mu} ( q_0  r _{\perp})
\over 2 ^{\mu+1} \pi \Gamma(\mu+1) },
\ee
where $\mu =  \alpha/2 -1$.

For distances smaller than the outer scale $r _{\perp} \ll q_0 $ and for 
$\alpha < 4$
this simplifies to
\be
A({\bf r}_{\perp})  = { B q_0^2 \over 2 \pi ( \alpha -2) }
\left( 1-  { \Gamma ( 2 - \alpha/2) \over \Gamma (\alpha/2) } 
\left( { q_0  r _{\perp} \over 2} \right)^{ \alpha -2} \right) 
\mbox{ $1/q_1 \ll r_{\perp} \ll 1/q_0$ and $\alpha < 4$}.
\ee
Using the definition of $B$ (Eq. \ref{coeffB}) we find
\be
A(r _{\perp}) =
\frac{4\sqrt{\pi }}{q_{0}}\delta ^{2}\frac{\Gamma (\frac{\alpha }{2})}{\Gamma (\frac{\alpha }{2}-\frac{3}{2})}
\frac{1}{\alpha -2}\left[ 1-\frac{\Gamma (2-\alpha /2)}{\Gamma (\alpha /2)}(\frac{q_{0}\rho }{2})^{\alpha -2}\right]
\approx 10 {\rm \, cm}.
\ee

On the other hand, 
for scales much smaller than the inner scale  $r _{\perp} \ll 1 /q_1$ 
\be
A({\bf r}_{\perp}) _{   r _{\perp} \ll 1 /q_1 } = 
{ B q_0^2 \over 2 \pi ( \alpha -2) } 
\left( 1- { 1\over 4}
 q_0^{ \alpha -2} q_1^{ 4 -  \alpha} r _{\perp} ^2 ( \alpha -2) \Gamma( { 4 -  \alpha \over 2} )
\right)
\mbox{ $ r_{\perp}  \ll  1/q_1 $}.
\ee
So that the typical length scale
\( z_{A} \) associated with A is
\[
z_{A}=\frac{\lambda ^{2}}{A(0)}\sim 100cm.\]

The 2-D structure functions of the index fluctuations
are 
\ba
&
D (r_{\perp})=
{ B q_0^2 \over \pi ( \alpha -2) } { \Gamma ( 2 - \alpha/2) \over \Gamma (\alpha/2) } 
\left( { q_0  r _{\perp} \over 2} \right)^{ \alpha -2} 
& \mbox{   $  1 /q_1 \ll  r _{\perp} $} 
\nonumber \\
&
D (r_{\perp})=
{ B q_0^2 \over 4 \pi } \Gamma( { 4 -  \alpha \over 2} ) q_0^{ \alpha -2} q_1^{ 4 -  \alpha} r _{\perp} ^2 
& \mbox{   $    r _{\perp} \ll  1 /q_1 $}
\label{structfunc}
\ea
Using the definition of B (Eq. \ref{coeffB}) we find for $  1 /q_1 \ll  r _{\perp} $
\be
D  (r_{\perp})=
\frac{8\sqrt{\pi }}{q_{0}}\frac{\delta ^{2}}{\alpha -2}\frac{\Gamma (2-\alpha /2)}{\Gamma
(\frac{\alpha }{2}-\frac{3}{2})}\left( \frac{q_{0}\rho }{2}\right) ^{\alpha -2}.
\ee

The condition $ D (r_{\perp})_S=1 $ defines $\rho_c $, the correlation length:
\be
\rho_c = {2 \over q_0} 
\left[ { 8 \pi \over B D k^2 q_0^2}  {\Gamma(\alpha/2) (\alpha-2) \over \Gamma(2-\alpha/2) } \right]^{1 /(\alpha-2)} =
\frac{2}{q_{0}}\left[ \frac{q_{0}}{Dk^{2}\sqrt{\pi }}\frac{\alpha -2}{\delta ^{2}}
\frac{\Gamma (\frac{\alpha }{2} - 
\frac{3}{2})}{\Gamma (2-\frac{\alpha }{2})}\right] ^{\frac{1}{\alpha -2}} \approx 10^{4} {\rm \, cm}.
\label{rho}
\ee
Note that in our case the correlation length is smaller than the Fresnel radius
$
{
\rho_c  \leq r_f} $.

The other  characteristic scale  of $D_S$, which determines the small scale $ r _{\perp} \ll 1 /q_1$ behavior,  is
\be
{1\over a_2^2} = \left| {\partial D_S  (r_{\perp}) \over \partial  r_{\perp}^2 }  
\right| _{r_{\perp} =0} \approx  \left( { 8 \pi \over B q_1^4 \Gamma( { 4 - \alpha \over 2}) }  \right)^{1/2}
\left( { q_1 \over q_0} \right)^{ \alpha/2}  \approx 10^{6} {\rm \, cm}.
\ee
Thus the structure function is almost constant for $ r _{\perp} \ll 1 /q_1$ and
has a typical correlation length $\rho_c $ given by Eq. (\ref{rho}).

As the waves propagate through a turbulent medium, the angular spectrum will tend to become Gaussian.
This occurs at (Lee \& Jokipi 1975a)
\be
z_G = { \lambda^2 \over B  q_1^2} \left( { q_1 \over q_0} \right)^{\alpha} 
{ \Gamma( 3- \alpha /2) \over \Gamma(2 - \alpha /2) } \approx 10^9 {\rm \, cm}.
\ee
This is much larger than the radius of the light cylinder, so that
the angular spectrum does not become Gaussian in our considerations.

There are several typical scattering angles associated with the power law distribution. 
First, there is the mean square angle:
\be
\left\langle \theta ^{2} \right\rangle = - {1\over 2} D  \nabla_{\perp}^2 D(r_{\perp})\left|_{r_{\perp}=0}
\right| =  
{ B D  q_0^2 \over 4 \pi } \Gamma( { 4 -  \alpha \over 2} ) q_0^{ \alpha -2} q_1^{ 4 -  \alpha}
 \sim 6  \times \, 10^{-6}.
\ee

 This is much smaller than the diffractive angle  given in Eq. (\ref{thetadiff}) for
small scale inhomogeneities 
and is comparable to the refractive scattering angle  Eq. (\ref{thetaref}) obtained using simpler
considerations. As we have mentioned, 
this is expected because our order-of-magnitude estimates for inhomogeneities
dominated by small scales  
 assumed that all of the power is concentrated at the diffractive scale. 
For the case of the power
law, the power in the density perturbations is attenuated by the large
phase space of long wavelengths refractive-type  perturbations.

The other  characteristic  angle $ \theta_c $, which  determines the   typical width of the angular power 
spectrum for a  power law distribution,
 is
defined by the correlation radius $\theta_c = {1\over k \rho_c } $:
\be
\theta_c = {  q_0 \over 2  k} 
{\left[ {  B D k^2 q_0^2  \over 8 \pi } { \Gamma(2-\alpha/2) \over \Gamma(\alpha/2) (\alpha-2) }
 \right]} ^{1 /(\alpha-2)} =
 { q_0 \over 2 k} {\left[ {\sqrt{ \pi} D k^2 {\delta}^2  \over  q_0 (\alpha-2)} 
{\Gamma(2-\alpha/2) \over \Gamma({\alpha - 3 \over 2} )}\right]} ^{1 / (\alpha-2)} \approx 10^ {-3} ,
  \mbox{ if $  1/q_1 \ll \rho_c 1/q_0$}.
\ee

The ratio of the two angles is
\be
{ \theta_c^2 \over \left\langle \theta ^{2} \right\rangle } \approx 0.2 
\mbox{ for $\alpha \approx 4$} 
\ee

The two angles scale differently with frequency: $ \left\langle \theta ^{2}\right\rangle $ 
is independent of frequency, while $ \theta_c \propto \nu ^ {  { 2 \over \alpha -2}} $ 
($\propto  \nu ^{ 1/5}$ for Kolmogorov spectrum).
They are just different measures of the angular spectrum;
 in principle, the two angles can serve to distinguish between Gaussian and power-law spectra of the 
density perturbations.

\subsection{ Temporal broadening and Decorrelation}

In the absence of dispersive decorrelation,
the intensity of the initial delta-pulse is a convolution
of refractive and diffractive contributions only:
\be
P(z,t)=P_D(z,t) \ast P_R(z,t).
\ee
The refractive Green function $P_R(z,t)$ has a symmetrical Gausian form (\cite{LJII}):
\be 
P_R(z,t)= { 2ck^2 \over \sqrt{2 \pi A(0) D}} \exp \left( {-2 c^2 k^4 (t-z/v_g)^2 \over A(0) D} \right) 
\ee
with characteristic refractive decorrelation time and bandwidth 
\ba
&&
\tau_R = \frac{1}{c}\left[ \frac{A(0)D}{2}\right] ^{1/2}= 10^{-5} {\rm \, sec}
\nn
\Delta \nu_R = { 1 \over 2 \pi \tau_R}  \approx 2  \times \, 10^4 {\rm \, Hz}
\ea
The diffractive Green function $P_D(z,t)$ has an approximately exponential shape 
with  a characteristic time of
\be
\tau_D ={ D \theta_c^2 \over 2 c}  \approx 10^{-7} {\rm \, sec},
\label{difftime}
\ee
and an  associated decorrelation bandwidth of
\be
\Delta \nu_D ={ 1 \over 2 \pi \tau_D }  \approx 2  \times \, 10^6 {\rm \, Hz}.
\ee
In case of power law spectrum
the size of the diffraction pattern  at Earth will be 
  $b \approx \lambda / \theta_{diff} \approx 10 ^{17} {\rm \, cm}$, where
\be 
\theta_{diff} \approx \theta_D {R_{LC} \over L} \approx 10^{-15} 
\ee
 is the visible size of the
scattering spot. The motion of this diffraction pattern due to the rotation
of the pulsar will produce variations on  a diffractive scintillations decorrelation  time scale of 
\be 
\tau_ {diff} = {b\over L} P  \approx {\lambda \over c \theta_D} \approx 10^{-10} {\rm \, sec},
\ee
with associated decorrelation bandwidth
\be
\Delta \nu_{diff} = {1 \over 2 \pi \tau_{diff}} \approx  10^{9} {\rm \, Hz}.
\label{decorband1}
\ee
These is a very larger decorrelation bandwidth. It is  of the order of the
observed frequency. Generally, the typical diffractive 
effects are weaker in the case of the power law spectrum of density inhomogeneities.
This is due to the fact that in case of power law spectrum
there is smaller power on the diffractive scales (given the fixed total power)
than in the case of single size inhomogeneities.
This makes  the diffractive 
effects  due to power law weaker and less
likely to be observed.

There though a possibility to distinguish power law from other types
of density spectra.
In case of power law distribution of inhomogeneities
both $\tau_D$  and $\delta \nu_D $ are weakly dependent on the frequency
($\tau_D \propto \nu ^{2/5} $) because of the frequency dependence of $\theta_c$. 
In contrast, if we have a Gaussian spectrum of density fluctuations, 
$\tau_R $  would be proportional  to $< \theta>$ and independent
of frequency.
 This again may possibly serve as a test for the type of power spectra:
for a Gaussian spectrum, the  diffractive decorrelation time $\tau_D$ should be independent
of frequency; while for a power law spectrum, $\tau_D  \propto \nu ^{2 (4- \alpha)/
(2-\alpha)} $.

\subsection{ Correlations of intensities}

The  correlation function of the wave intensities does not have an analytical
representation  (Rytov et al.). Asymptotic expressions are possible in the
 cases of weak or very strong scintillations. The division between these two
limits is based on whether $D( r_f)_S$ - the
value of the structure function of the phase correlations
at  the first Fresnel zone -
is $\ll 1$ or $\gg 1$.  This may be cast in the form of 
 the ratio of the first
Fresnel zone to the coherence radius $\sqrt{\lambda D}/\rho_c$:
scintillations are weak for $ \sqrt{\lambda D}/\rho_c \ll 1$, and
strong in the opposite case.

It is easy to see  using Eqns (\ref{structfunc}) that in our case 
 $D(r_f)_S  \approx 100 \gg 1$.
In this limit
 the scintillation becomes saturated with the scintillation index reaching 1.
According to Prokhorov et al. 1975,
there are two regimes for the spectrum of scintillations,
separated by $ q^{\ast} \approx 1/\sqrt{\lambda D} D(r_f)_S ^{ {\alpha -1 \over \alpha (\alpha+1) }} 
\approx 1/r_f$
(this is equivalent to the separation between diffractive
and refractive scintillation).
For $  q  \ll q^{\ast} \approx 1/r_f$ the modulations are weak, and the modulation spectrum
is $\propto q^4$ (Prokhorov  et al. 1975 (4.40)) - these are diffractive scintillations.
For $  q  \gg q^{\ast} $ - refractive scintillations -
  modulations are strong and saturated:  $m_z \approx 1$.

If $D(r_f)_S  >1 $ 
the scintillation index is  (Prokhorov  et al. 1975)
\be
m^2 \approx 1+ {\rm \, C} D(r_f)_S ^{ - { 2 \over \alpha (2 -\alpha ) }}
\ee
where ${\rm \, C} $ is of the order of unity.

Thus we conclude that strong saturated refractive scintillations are possible
inside pulsar magnetosphere.

\section{Observational Tests}

Here we summarize the predicted characteristics of the scattering  inside
the pulsar magnetosphere:
\begin{tabbing}
diffractive scattering angle \hskip 1 truein \= $ 10^{-1 } $ \\
diffractive scattering  time \> $10^{-4} {\rm \, sec}$ \\
diffractive decorrelation time \> $10^{-8  }  {\rm \, sec}$ \\
refractive  scattering angle  \>   $ 10^{-3  } $ \\
refractive  decorrelation time \> $10^{-4 } {\rm \, sec} $ \\
arrival time variations \> $10^{-4 } {\rm \, sec}$ \\
\end {tabbing}
All these quantities  are independent of frequency,
but  the  strength of scattering  increases with  
frequency. This frequency dependence of the strength of scattering  is opposite to
the interstellar propagation effects (which are weakest at larger frequencies).
Thus the weak magnetospheric effects should be most prominent  at high
frequencies in nearby pulsars.

In conclusion we discuss some observational facts that may possibly be attributed
to magnetospheric scattering and propose  future or follow up experiments to 
test the theory.

Some of the propagation effects have possibly been observed.   
The most interesting and decisive (in our opinion) 
observations of Sallmen et al. (1999) became known to us 
 when most of  the present work was  completed. Two results of this work strongly 
support our theory:  the  frequency independent jitter in the arrival  time,
of the order $100 \mu$sec; 
and the frequency  independent  spread of the multiple components of the
giant pulses, of tens of $\mu$sec,  with various scattering times  also of  the order of tens of $\mu$sec.
Using our results, these observations imply that 
the multiple structure of pulses  is due to the   multipath propagation 
inside the pulsar magnetosphere: it is frequency independent, but different rays
propagate different lengths, aquiring a range of scattering times.

Other observations that can be interpreted in favor of magnetospheric propagation include
the following.
Large  sizes of the emission region(Gwinn et al. 1997, Smirnova et al. 1996
and Cordes et al. 1983)  may be due to diffractive scattering.
Increase in the temporal broadening time $\tau_D \approx 10^{-8} {\rm \, sec}$
for nearby pulsars with dispersion measure $DM\leq 20$,
relative to that extrapolated from more distant pulsars
(Britton et al. 1998) may be due to diffractive  decorrelation inside the 
pulsar magnetosphere. This  $\tau_D $ is comparable to our prediction (Eq. \ref{difftime})
if this is due to diffractive decorrelation inside the magnetosphere.
Similarly, the predicted frequency independence of the decorrelation bandwidth (Eq. \ref{decorband})
 naturally explains the
results of Gwinn at al. (1999) - the  diffractive  broadening time $\tau_D \approx 10^{-8} {\rm \, sec}$
will  produce the observed  5 MHz bandwidth gives.
In the same manner the
unusual decorrelation bandwidth of the PSR 0950 (Kramer et al. 1999) 
of 8 MHz is close to the predicted due to the diffractive
scattering. If a  follow up observation at a different frequency would show the
same decorrelation bandwidth it will a strong argument in favor of the
magnetospheric scattering.
The enhanced  intensity fluctuations (if compared with extrapolations from lower frequencies)
at very high
frequencies (30 GHz) found by Kramer et al. (1997) in several low DM pulsars
maybe due to the  scintillations inside
the pulsar magnetosphere.

The observations of B0950+08 at very low frequencies of 60 and 102 MHz by
Smirnova \& Shabanova (1992) which showed frequency independent narrowband variation of the pulse profile at
both frequencies, with a characteristic bandwidth of 30 to 40 kHz
may be explained as been due to  refractive-type events with a typical refractive
time $\tau_R \approx 3  \times \, 10^{-4} {\rm \, sec}$ (Eq. \ref{nuR})

Experiments to detect effects
 of wave propagation inside the pulsar magnetosphere should use high frequencies for observation, and concentrate on nearby
pulsars with low dispersion measure. 
Possible  experiments  will include a search for nondispersive (frequency independent) effects such as a
time delay (as large as tens of microseconds) in the pulse arrivals, 
a  diffractive   decorrelation bandwidth of the order of $10$ MHz, 
 and microstructure periodicities (of the order of tens of microseconds) due to refractive scattering.
Nearby strong pulsar,  like PSR 0950, are best  candidate for searches for magnetospheric
effects.

Other  
 possible experiment that may be used to search for diffractive scattering 
inside the pulsar exploits interstellar scattering effects.  Consider  a thin  interstellar screen
which 
scatters pulsar radiation. The size of a patch on the screen is, approximately,
the width of the pulsar diagram $\Delta W$ times the distance from the pulsar to the screen $z$. 
If there were  considerable diffractive scattering inside the magnetosphere,
the size of the coherent patch would be $ \lambda z / (\theta _s R_{LC} ) \ll z \Delta W$. 
A smaller size of the coherent patch would change the  observed properties of the diffraction pattern
observed at Earth.
Also,
future high temporal resolution studies of pulsars  (at time scales of nanoseconds)
should provide  more information on the weakest scattering events and possibly distinguish
between interstellar and magnetospheric events.

\acknowledgements

We would like to thank Carl Gwinn for  numerous discussions and his  valuable comments. 
AP thanks CITA for its hospitality during his stay in Toronto. This research was partially 
supported by NSERC.

\appendix
\section{Dispersion relation for streaming pair plasma}
\label{modes}

If the average velocities of the electron and positrons of the secondary plasma
are the same,
the normal modes of strongly  magnetized electron-positron plasma
consist of three wave branches: extraordinary (X) and two coupled
 ordinary (O)
and Alfv\'{e}n branches.
Alfv\'{e}n mode cannot leave magnetosphere, while
X and O mode may leave magnetosphere.

For the  forward propagating waves in the pulsar frame we have
\begin{eqnarray}
&
\omega_X  = k
c\left( 1 - {  \gamma_p \omega_p^{ 2}
( 1-v_p \cos \theta /c)^2
\over \omega_B^2} \right) =
\left\{
\begin{array}{ll}
 k  c\left( 1-
 {  \omega_p^{ 2} \over 4 \gamma_p^3 \omega_B^2} \right), &
\mbox{ if $\theta \ll { 1 \over \gamma_p} $}\\
 k  c\left( 1- { 4 \gamma_p  \omega_p^{ 2}  \sin ^2 {\theta \over 2} \over \omega_B^2}
\right),
& \mbox{ if $   \theta \gg  { 1 \over \gamma_p} $ }
\end{array} \right.
&  \mbox{} \nonumber \\ \mbox{}
&
\omega_O =
 k
 c \left( 1 -
 {  \gamma_p \omega_p^{ 2} ( v_p/c - \cos  \theta ) ^2
\over  \omega_B^2}
+ { \omega_p^{ 2}
\,\sin^2\theta   \over
     \gamma_p^3    c^2\,k^{ 2} \, ( 1- v_p \cos \theta/c)^2
           }  \right)
=\left\{
\begin{array}{ll}
 kc \left( 1-
 {  \omega_p^{ 2} \over 4 \gamma_p^3 \omega_B^2} \right)
& \mbox{ if $ \theta \ll  \omega /\gamma_p^2 \omega_B$} \\
 kc \left( 1 + {\frac{ 4 \, \gamma_p{{\, \omega_p}^{ 2}
}\,{{\sin^2\theta  }}}
        {{c^2}\,{k^{ 2} }\,
 }}  \right)
& \mbox{ if $  \omega /\gamma_p^2 \omega_B \ll \theta \ll 1/\gamma_p $} \\
 kc \left( 1- {   4  \gamma_p \omega_p^{ 2}  \sin ^2 {\theta \over 2} \over  \omega_B^2} +
{ \omega_p^{ 2} \cot^2 { \theta \over 2} \over \gamma_p^3  k^2 c^2 } \right)
&  \mbox{ if $   \theta \gg  { 1 \over \gamma_p} $ }
\end{array} \right.
& \mbox{ if $ k  c \,\gg  \,\gamma_p  \omega_p  $ }
\mbox{} \nonumber \\ \mbox{}
& \omega_A  =
k  c \cos\theta
 \left(1 -
 { \, \omega_p^{ 2}\,
\over 4  \,  \gamma_p^3  \, \omega_B^2 }   \,
   - {\frac{{c^2}\,{k^{ 2} }\,{{\sin^2\theta  }}}
        {16\,\gamma_p {{\, \omega_p}^{ 2}}}}   \right)
=\left\{
\begin{array}{ll}
 kc  \cos\theta  \left( 1-
 {  \omega_p^{ 2} \over 4 \gamma_p^3 \omega_B^2} \right)
 & \mbox{ if $ \theta \ll
{  \omega_p^2 \over  \gamma_p
 \omega_B \omega} $} \\
kc  \cos\theta \left( 1-  {c^2\,k^ 2 \, \sin^2\theta  \over
        4\,\gamma_p \, \omega_p^ 2} \right)
  & \mbox{ if $ \theta \gg {  \omega_p^2 \over  \gamma_p
 \omega_B \omega} $}
\end{array} \right.
& \mbox{ if $ k  c \,\ll  \,\gamma_p  \omega_p  $ }
\label{qw}
\end{eqnarray}


\begin{thebibliography}{}

\bibitem{Arons1983}
Arons J. 1983,
\apj, 266, 215

\bibitem{Bhat1}
Bhat N.D.R., Gupta Y. \& Rao A.P. 1999,
ApJ, 514, 249


\bibitem{Bhat2}
Bhat, N.D.R., Gupta, Y., \& Rao, A.P. 1998, ApJ, 500, 262 

\bibitem[Blandford and Narayan 1985]{Blandford&Narayan}
Blandford R. and Narayan, R. 1985,
Mon. Not. R. astr. Soc., 213, 591


\bibitem{Britton}
Britton M.C., Gwinn C.R. \& Ojeda M.J. 1998,
ApJ, 501, L101


\bibitem[Cordes, Weisberg \& Boriakoff 1985]{cor85} Cordes, J.M., Weisberg, J.M., \& Boriakoff, V. 1985, ApJ, 288, 221

\bibitem{Hankins}
Hankins, T.H., Moffett, D.A. 1998, BAAS, 192, 57.02 

\bibitem{Gupta}
Gupta, Y., Rickett, B.J., \& Lyne, A.G. 1994, MNRAS, 269, 1035 

\bibitem[Gwinn et al. 1997]{gwi97}
Gwinn, C.R. et al. 1997, ApJ, 483, L53 

\bibitem{comments}
Gwinn, Hirano, Britton (1999) personal comm. 

\bibitem{Jenet98}
Jenet, F. A., Anderson, S. B., Kaspi, V. M., Prince, T. A., Unwin, S. C. 1998,
ApJ, 498, 365


\bibitem[Jenet et al. 1997]{JenetCook}
Jenet F.A. et al. 1997,
Pub. Ast. Soc. Pac., 109, 707

\bibitem[Kramer  et al. 1997]{Kramer97}
Kramer M., Xilouris K.M., Rickett B. 1997,
A\&A, 321, 513

\bibitem{Kramer}
{Kramer}, M. et al. 1999,
\apj, 520, 324

\bibitem[Lee and Jokipii  1975a]{LJI}
Lee L.C. and Jokipii J.R. 1975a
ApJ, 196, 695

\bibitem[Lee and Jokipii  1975b]{LJII}
Lee L.C. and Jokipii J.R. 1975b
ApJ, 201, 532

\bibitem[Lee and Jokipii  1975c]{LJIII}
Lee L.C. and Jokipii J.R. 1975c
ApJ, 202, 439

\bibitem[Melrose 1995  ]{Melrose-DB}
Melrose D.B. 1995, J. Astroph. Astron., 16, 137


\bibitem{Prokhorov}
Prokhorov A.M., Bunkin F.V., Gochelashvili K.S. \& Shishov V.I. 1975,
Sov. Phys. Usp., 17, 826

\bibitem{Rickett75}
Rickett B.J. 1975,
Ann. Rev. Astron. Astroph, 15, 479

\bibitem{Rickett1}
Rickett, B.J., Coles, W.A., \& Markkanen, J. 1999, submitted to MNRAS

\bibitem{Rytov}
Rytov S.M., Kravtsov Yu.A., Tatarskii V.I. 1989,
{\it Principles of statistical radiophysics},
Berlin ; New York : Springer-Verlag

\bibitem{Sallmen}
{Backer}, D. C., {Hankins},T. H., 
        {Moffett}, D. \& {Lundgren}, S. 1999, 
\apj, 517,460

\bibitem[Smirnova et al. 1996]{smi96} Smirnova, T. V., Shishov, V. I., Malofeev, V. M. 1996, ApJ, 462, 289

\bibitem[Weatherall]{Weatherall}
{Weatherall} J. C. 1997,
\apj, 483, 402

\bibitem[Wolszczan \& Cordes 1987]{wol87} Wolszczan, A. \& Cordes, J.M. 1987, ApJ, 320, L35

\bibitem[Zhelezniakov 1996]{Zhelezniakov1996}
Zhelezniakov, V.V. 1996,
{\it Radiation in astrophysical plasmas },
Dordrecht ; Boston : Kluwer

\end{thebibliography}
\end{document}